\documentclass[aps,prl,preprint,superscriptaddress]{revtex4}

\usepackage{color}
\usepackage{graphicx}
\usepackage{bm}
\usepackage{amsmath}
\usepackage{amssymb}
\usepackage{hyperref}
	
\begin{document}
	
\title{Size effect on the spontaneous coalescence of nanowires}

\author{Zhenyan Wu}
\affiliation{Department of Physics, Guangxi University, Nanning 530004, P. R. China.}

\author{X. Yang}
\affiliation{Department of Physics, Guangxi University, Nanning 530004, P. R. China.}

\author{Zhao Wang}
\email{zw@gxu.edu.cn}
\affiliation{Department of Physics, Guangxi University, Nanning 530004, P. R. China.}
		
\begin{abstract}
This paper investigates the size effect on the coalescence process of contacting nanoparticles. It is revealed by molecular dynamics that the nanometer-sized surface curvature coupled with the effective melting temperature exhibits a strong influence on the atom diffusion at the interface, and is therefore critical to the coalescence time. This effect is particularly pronouncing for surface curvatures below 20 nm. A phenomenological model is derived from the melting-point reduction approach to describe the kinetic process of nanowire coalescence and is validated against a variety of simulation datasets. The quantitative correlation between the sample size, the sintering temperature and the contact morphology evolution is demonstrated.
\end{abstract}
		
\maketitle

\section{Introductions}
The spontaneous coalescence is critical for the self-assembly of nanomaterials. It is known to be strongly correlated with the surface atom diffusion, which manifests when the material size goes down below tens of nanometers. Experiments demonstrated significant surface diffusion taking place even at room temperature at sub-$10\;\mathrm{nm}$ lengthscale \cite{Akande2010}. Similar behaviors were reported by experiments of the cold-welding of gold nanowires (NWs) \cite{Lu2010} and NPs \cite{Wagle2015,Sun2014}, and of self-assembly of nanoparticle (NP) aggregates \cite{Klajn2007}. Understanding the size-effect on the mass-diffusion process at nanoscale solid interfaces is crucial for developing self-assembly technologies of nanostructures, which hold promise for a wide range of applications \cite{Liu2013}.

Recently, it is reported by simulations that the coalescence of NPs can start without the thermal activation \cite{LiQb2016}, and that the NP size and the sintering temperature exhibit significant effects on the densification of the sintered nanoparticals \cite{Yang2018}. Lu \textit{et al.} demonstrate that single crystalline gold NWs with diameters between 3 and 10 nm can be cold-welded together within seconds by mechanical contacts alone under low applied pressures \cite{Lu2010}. Su \textit{et al.} report that the gold NPs can be self-assembled at silver NW junctions and nanogaps by rod-coating \cite{Su2019}. Sabelfeld and Kablukova develop a stochastic model of the growth of an ensemble of GaN NWs to include the coalescence caused by bundling \cite{Sabelfeld2018}. The surface chemistry and the chemical nature of the material are found to strongly influence the process of the cold welding of nano-objects \cite{Wagle2015}.

The coalescence process is driven by a natural need to minimize the surface chemical potential. Hence the mass diffusion at sub-10nm-curvature surface is mainly driven by a dramatic increase in the surface energy \cite{Jose-Yacaman2005}. The theory of macroscopic thermal grooving describes the evolving shape of particles or wires in coalescence by considering both evaporation and surface diffusion mechanisms \cite{Mullins1957}. Despite of successful applications in interpreting a number of experimental measurements, this model suffers from problems caused by its assumptions ignoring the atomistic details of the system. Meanwhile, atomistic simulations have intensively been used to study the sintering \cite{Cheng2013,Buesser2011,Koparde2005} and coalescence \cite{Wang2016,Lim2009,Hawa2005,Li2016a,Guevara-Chapa2014,Grammatikopoulos2014} processes in nanomaterial synthesize. Notably, the mechanisms of melting temperature variation and phase transformation have been quantified by Koparde and Cummings \cite{Koparde2008b,Koparde2008a}. However, little is known up to date, about the combined roles of the sample size and temperature in the kinetic process of NP coalescence.

To this end, here we simulate the spontaneous coalescence of two contacting NWs using molecular dynamics (MD) \cite{Wang2018,Wang2011,Qi2018,Wang2009c,Guo2015}. The essential role of the surface curvature coupled with the thermal effect in the coalescence process is demonstrated. Moreover, the simulation results are used as inputs for developing a phenomenological model. Unlike previous models, the present model takes the curvature-dependent surface diffusivity into account, and is able to predict the coalescence time as a function of NW size and the temperature.

\section{Methods}
\begin{figure}[htp]
\centerline{\includegraphics[width=11cm]{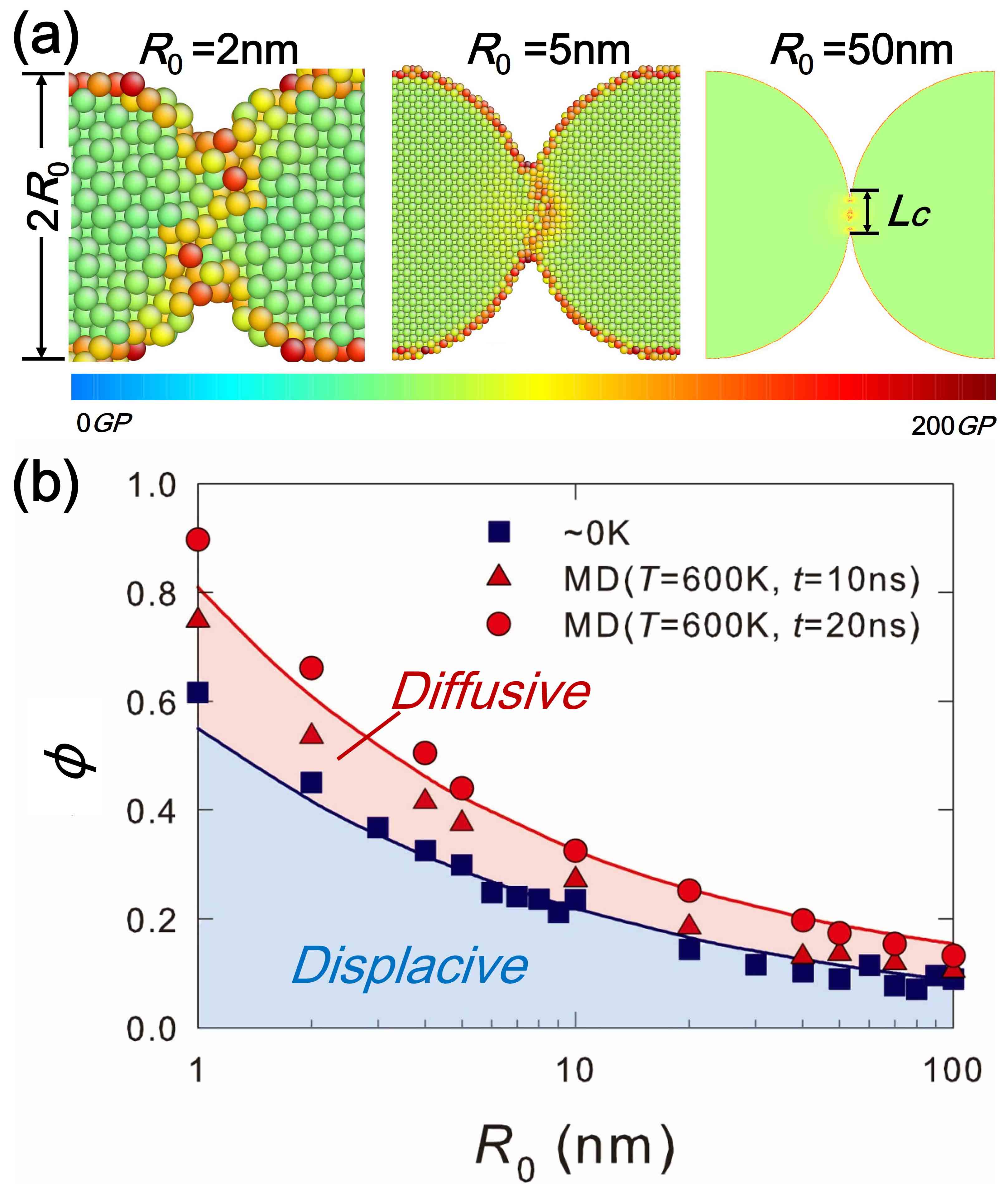}}
\caption{\label{fig:1}
(a) Snapshots of the simulated contact between a pair of Al NWs (cross-section view along [100] crystallographic direction). The color scale corresponds to the von-Mises stress distribution. (b) Contact area ratio $\phi=A/A_{0}$ as a function of $R_{0}$. The squares represent the ground-state ($0\;\mathrm{K}$ limit) results, while the circles and triangles stand for those of MD for simulation time of $10$ and $20\;\mathrm{ns}$, respectively. The contact area $A$ is computed by defining an inter-particle spacing cutoff of $0.286\;\mathrm{nm}$ as the equilibrium inter-atomic distance and $A_{0}=2 w R_{0}$  where $w$ is the thickness.}
\end{figure}

We start by simulating the contact between two curved aluminum NW surfaces at zero external load. This set-up mimics a NW cold-welding experiment \cite{Lu2010}, as shown in Fig.\ref{fig:1}(a). The simulations are performed in a two-dimensional plane-stress configuration, with a periodicity $w \approx 1.62 \,$nm in the direction perpendicular to the cross section plane. An embedded atom method (EAM) is used to describe the potential energy $\varepsilon$ of the interaction between the Al atoms,

\begin{equation}
\label{eq0}
\varepsilon(r)=[\frac{V_{0}}{(b_{2}-b_{1})}(\frac{b_{2}}{z^{b_{1}}}-\frac{b_{1}}{z^{b_{2}}})+\delta]\Psi(\frac{r-r_{c}}{h}),
\end{equation}

\noindent where $r_{c}$ is the cutoff distance, $z=r/r^{'}$, and $b_{1}$, $b_{2}$, $\delta$, $V_{0}$, $h$ and $r^{'}$ are fitting parameters, $\Psi(x)$ is a cutoff function. The parameterization of this EAM force field is provided in Ref.\onlinecite{Zope2003}. We use a Nos$\acute{e}$-Hoover thermostat at a time step of $0.5\;\mathrm{fs}$ to simulate the shape evolution of the contact \cite{Yang2015a}. The system temperature is controlled to be relatively high for letting the system reach an equilibrium state in the time scale accessible to MD, i.e. in the order of nanoseconds, since the atomistic diffusion can be strongly accelerated at temperatures close to the bulk melting point \cite{Guo2015,Tianou2017}. Note, that surface diffusion is observable even at room temperature in experiments \cite{Lu2010}, since the experimental time scale is typically $10-13$ orders of magnitudes larger than that of classical MD \cite{Li054103}.

\section{Results and Discussions}
The ratio between the effective contact area and apparent one can be greatly enhanced by decreasing the surface curvature to nanometer-scale \cite{Guo2015}. This size effect becomes most pronounced for tip radii below $10\;\mathrm{nm}$, as shown in Fig.\ref{fig:1}(b). Similar to previously reported experimental \cite{Lu2010,Bay1983,ChengL2019} and computational \cite{Pereira2011} results, the reconstruction of the cubic lattice with very few defects is observed. The displacive plasticity is simply related to the electrostatic nature of the inter-atomic force, as a sharp tip contains a larger fraction of surface atoms that are exposed to the atomistic attractive force of the adjoining surface, which decreases rapidly with increasing separation distance and tends to vanish after several nanometers. This is consistent with the experimentally observed size effect on the contact between NPs and NWs \cite{Tang2002}, and is strongly correlated with the inverse contact scaling in biological adhesive systems \cite{Autumn2000,Lee2007}. For instance, a ten-fold-increase is found for an $R_{0}=1$ nm contact at $600\;\mathrm{K}$.

\begin{figure}[htp]
\centerline{\includegraphics[width=11cm]{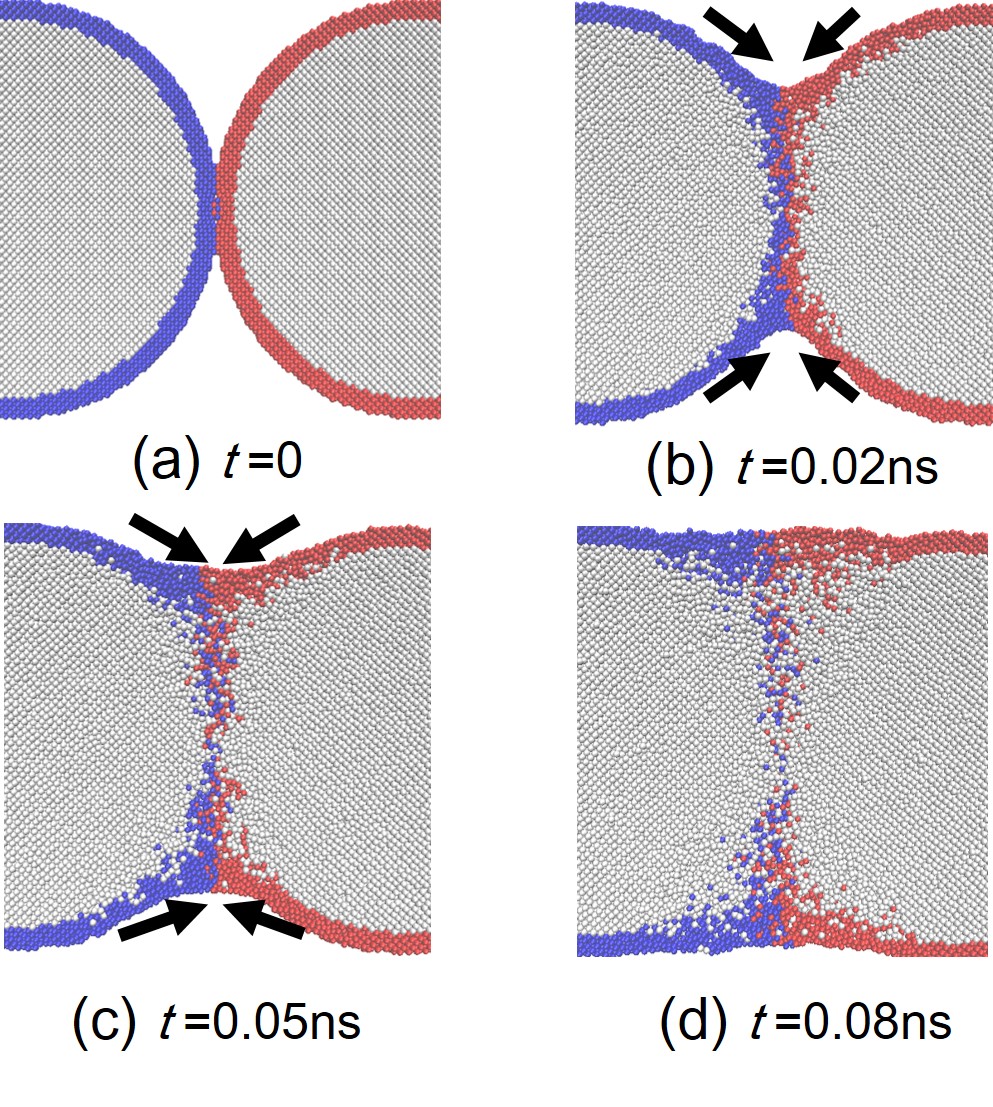}}
\caption{\label{fig:2}
(a-d) Snapshots of the simulation on adhesion of two NWs ($R_{0}=10\;\mathrm{nm}$) with fixed boundaries at $900\;\mathrm{K}$. The arrows represent the direction of surface atom flow. The surface atoms are labeled with different colors.}
\end{figure}

Results are obtained at different temperatures. The ``zero-K'' results [squares in Fig.\ref{fig:1}(b)] are from molecular mechanics \cite{Wang2009,Wang2007a,Wang2007} which do not include the thermal effects, and thus only represent the time-independent (so-called displacive) contribution to the contact area. Comparing the two sets of $600\;\mathrm{K}$ data [triangles and circles in Fig.\ref{fig:1}(b)] that were obtained by MD with different simulation time, we see that the finite-temperature contribution to the contact area is time-dependent. This time-dependency is strongly correlated with the surface atom diffusion, as shown in Fig.\ref{fig:2} for a rigid-boundary set-up. The colors contrast the surface and in-body atoms, by which we clearly see that the increase in the contact area is mainly contributed from immigrated surface atoms. We see that, starting from the initial displacive contact [Fig.\ref{fig:2}(a)], the surface atoms of the two contacting bodies diffuse into the neck region [Fig.\ref{fig:2}(b,c)] until the curvature radius tends to be uniform along the surface [Fig.\ref{fig:2}(d)]. This is consistent with microscopy experiment observations \cite{Laza2013,Honey2015}.

\begin{figure}[htp]
\centerline{\includegraphics[width=11cm]{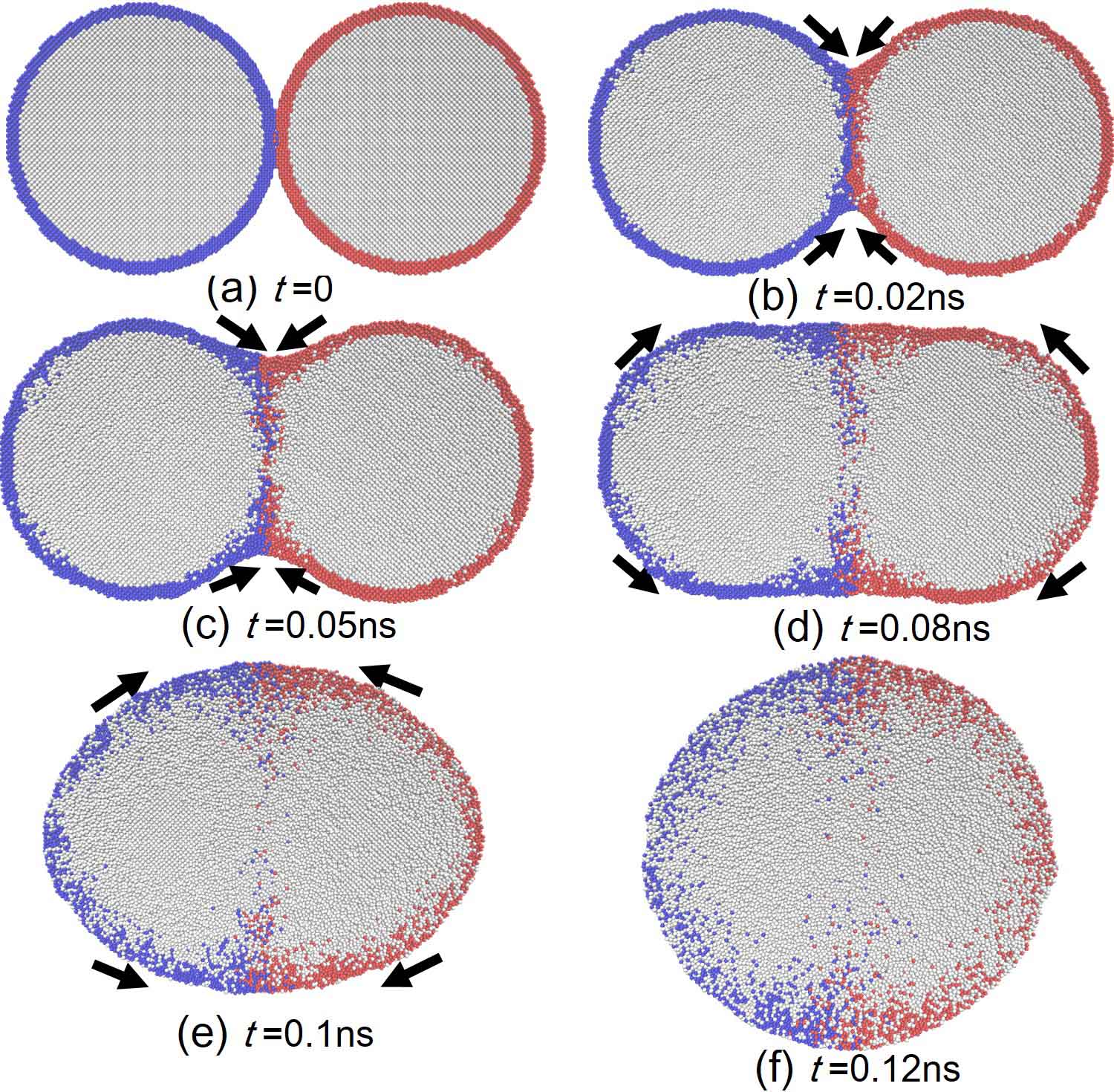}}
\caption{\label{fig:3}
(a-f) Snapshots of the coalescence of two NWs ($R_{0}=10\;\mathrm{nm}$) with free boundaries at $900\;\mathrm{K}$ from cross-section view. The arrows represent the direction of surface atom flow. The surface atoms are labeled with different colors.}
\end{figure}

Another set of simulation is performed for NWs with free boundaries. It can be seen in Fig.\ref{fig:3}(a-d) that the neck is filled with  diffused atoms with time, similar to that in the fixed-boundary case shown in Fig.\ref{fig:2}. However, the wire shape keeps evolving until the formation of a new wire, as shown in Fig.\ref{fig:3}(e-f). This observation is consistent with the field-emission TEM results reported by Cheng \textit{et al.}, which show the e-beam-induced coalescence between NWs and NPs takes place by the fast, massive atom transportation near their contact surface region \cite{ChengL2018,ChengL2019}.

To quantify the diffusion behavior of surface atoms, we consider a phenomenological description derivable from the concept of curvature-dependent surface potential energy \cite{Mullins1957},

\begin{equation}
\label{eq4}
\Delta \phi =\frac{\beta}{A_{0}}\int_{0}^{t} \Delta D dt,
\end{equation}

\noindent where $\beta$ is a positive constant and $\Delta D$ is the difference in the diffusivity of the atoms at the free surface and that of the atoms at the neck region. In contrast to the original model of Mullins \cite{Mullins1957}, we consider $D$ is no longer constant but changes with $R$. This introduces the concept of curvature-dependent diffusivity of nano-crystals. This is based on the melting-point reduction approach \cite{Jiang04,Guisbiers08} that was used to study atom diffusion in sintering of silica-encapsulated Au \cite{DickDZM02} and Au-Ag NPs \cite{ShibataBZMVG02}. This allows describing the effective diffusivity of surface atoms as a function of the surface curvature radius and melting point shift,

\begin{equation}
\label{eq1}
 D=D_{0}\exp\left[-\frac{C T_{m,\infty}}{k_{\rm B}T}\left(1-\frac{\alpha}{2R}\right)\right].
\end{equation}

This equation is accompanied with the well-established Gibbs-Thomson equation \cite{BuffatB76,ZhangESOKLWGA00}, by which the melting point is approximately a function of the surface curvature radius $R$ of small crystals,

\begin{equation}
\label{eq2}
T_{m}(R)=T_{m,\infty}\left(1-\frac{\alpha}{2R} \right),
\end{equation}

\noindent where $T_{m,\infty}$ is the bulk thermodynamic melting point, $k_{\rm B}$ is the Boltzmann constant, both $D_{0}$ and $C$ are temperature- and size-independent positive constants. The shape parameter $\alpha$ can be obtained by considering the effect of size and shape on the bulk melting temperature,

\begin{equation}
\label{eq3}
\alpha=\frac{4\upsilon_{sl}}{H_{{f}}\rho_{s}},
\end{equation}

\noindent where $\upsilon_{sl}$ is the solid-liquid interface energy, $H_{f}$ is the bulk enthalpy of fusion and $\rho_{s}$ is the solid state density.

\begin{figure}[htp]
\centerline{\includegraphics[width=11cm]{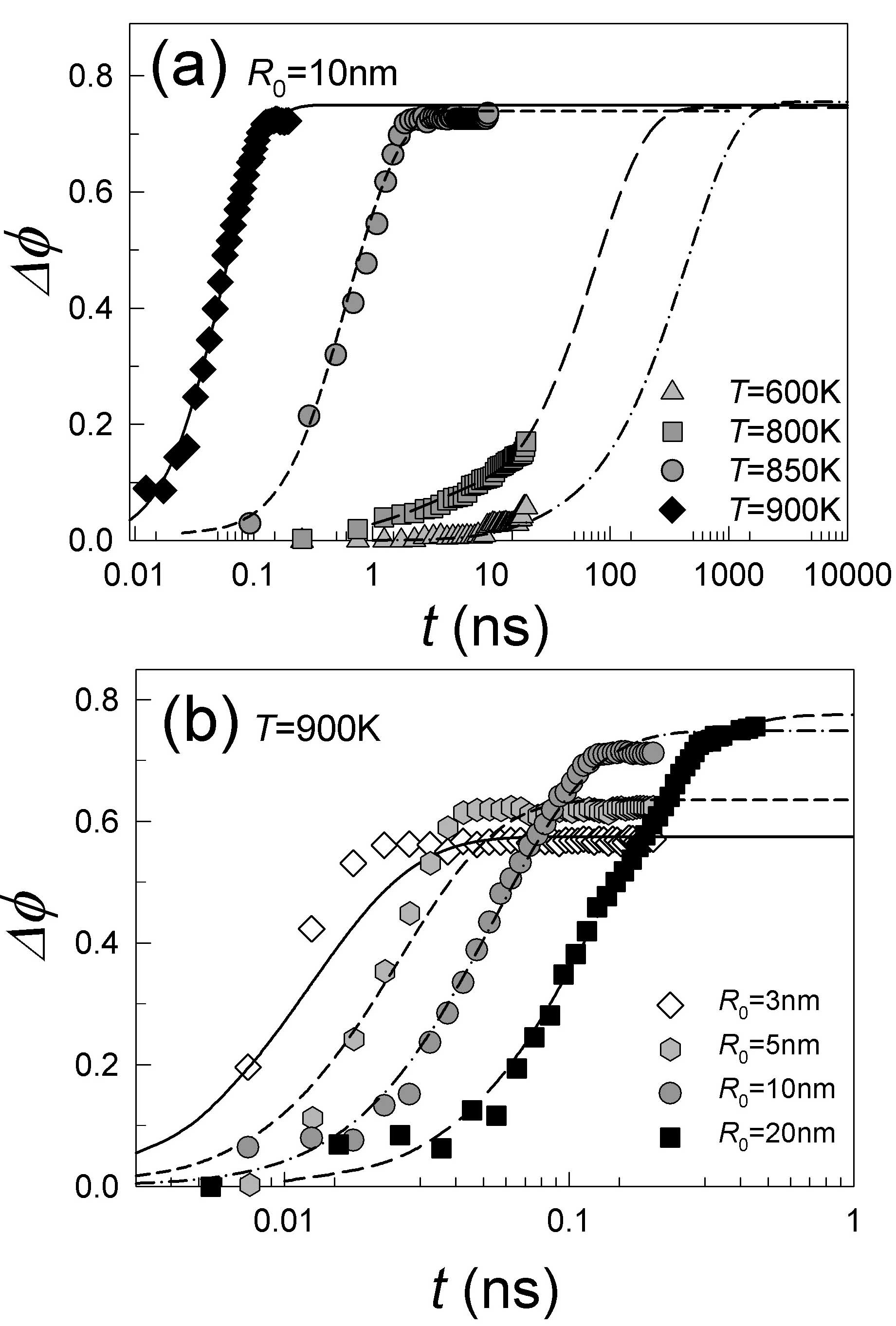}}
\caption{\label{fig:4}
Change in contact area ratio $\Delta \phi$ \textit{versus} time for a pair of NWs at different temperatures (a), and for those of different sizes at a given temperature (b) with fixed boundaries. The symbols stand for the simulation data and the curves represent the results obtained by Eq.\ref{eq8}.}
\end{figure}

Numerical analyses show that the time-dependent diffusivity difference $\Delta D$ can be approximated by an exponential-decay function of time $t$, as shown in the supplementary material,

\begin{equation}
\label{eq7}
\Delta D = D_{0}\left\lbrace  \exp \left[ -\frac{C T_{m}(R_{0})}{k_\textbf{B} T}\right] - \exp \left[-\frac{C T_{m,\infty}} {k_\textbf{B} T}\right]\right\rbrace \exp(- \gamma t),
\end{equation}

\noindent where the effective melting temperature of the surface atoms at the neck region is assumed to be $T_{m,\infty}$, and the parameter $\gamma$ denotes the decay rate of $\Delta D$. This yields

\begin{equation}
\label{eq8}
\Delta \phi =\frac{\beta}{A_{0}}\int_{0}^{t} D_{0}\left\lbrace  \exp \left[ -\frac{C T_{m}(R_{0})}{k_\textbf{B} T}\right]-\exp \left[-\frac{C T_{m,\infty}} {k_\textbf{B} T}\right]\right\rbrace \exp(- \gamma t) dt.
\end{equation}

\noindent where the values of $\gamma$ and $\beta$ are obtained by fitting to simulation results. The parameter values used in this work are provided in the supplementary material.

The simulated evolution of the contact morphology can be well predicted by Eq.\ref{eq8} for different sizes and temperatures, as shown in Fig.\ref{fig:4}. It can be seen that the contact area reaches a limit with different adhesion velocities at different temperatures for a given wire radius [Fig.\ref{fig:4}(a)]. For instance, for $R_{0}=10$nm, the time required for $\phi$ to saturate at $600\;\mathrm{K}$ is over four orders of magnitude longer than that at $900\;\mathrm{K}$. This observation is qualitatively consistent with the results reported  by Cheng \textit{et al.} \cite{Cheng2010a,Cheng2010b}. We also see that the small contact exhibits higher adhesion velocity than the large ones [Fig.\ref{fig:4}(b)]. The difference between the prediction by Eq.\ref{eq8} and MD becomes more significant when $R_{0}$ decreases below $5\;\mathrm{nm}$. This may be due to the effect of surface roughness \cite{Luan2005b}.

\section{Conclusion}
In conclusion, we have demonstrated by MD and an analytical model that the nanometer-sized surface curvature coupled with the effective melting temperature is critical to  the NP coalescence.  We develop a phenomenological model to predict quantitatively the NW morphology evolution as a function of the NW size and the coalescence temperature, by taking into account the curvature-dependent surface diffusivity. These results have strong implications to our understanding of the mass diffusion at sub-$10\;\mathrm{nm}$ lengthscale. We remark that classical MD with a typical time step of $1.0\;\mathrm{fs}$ could be limiting to adequately capture real material processes taking place on much longer time scale. Also the simplistic way of modeling the surface diffusion used in this work may not realistically represent the effect of local environment, such as surface passivation (e.g. oxygen, hydrogen, liquid solutions and other adsorbates) in real experiments. One can however envision implementing the Monte Carlo \cite{Henkelman2001} or diffusive MD \cite{Li054103} simulations in light of the above-introduced kinetic model. The combination of atomistic simulations and continuum contact theories should be further applicable to a wider range of surface types, and is expected to provide useful guidelines for experimentalists working on welding and self-assembly of nano-objects, that are otherwise mostly limited by a time- and effort-consuming trial-and-error procedure.

\end{document}